# Aldebaran's angular diameter: How well do we know it?*

A. Richichi[1] and V. Roccatagliata[1,2]

[1] European Southern Observatory, Karl-Schwarzschildstr. 2, 85748 Garching bei München, Germany
   e-mail: `arichich@eso.org`
[2] Dipartimento di Astronomia, Università di Padova, Vicolo dell'Osservatorio 2, 35122 Padova, Italy

**Abstract.** The bright, well-known K5 giant Aldebaran, $\alpha$ Tau, is probably the star with the largest number of direct angular diameter determinations, achieved over a long time by several authors using various techniques. In spite of this wealth of data, or perhaps as a direct result of it, there is not a very good agreement on a single angular diameter value. This is particularly unsettling if one considers that Aldebaran is also used as a primary calibrator for some angular resolution methods, notably for optical and infrared long baseline interferometry. Directly connected to Aldebaran's angular diameter and its uncertainties is its effective temperature, which also has been used for several empirical calibrations. Among the proposed explanations for the elusiveness of an accurate determination of the angular diameter of Aldebaran are the possibility of temporal variations as well as a possible dependence of the angular diameter on the wavelength. We present here a few, very accurate new determinations obtained by means of lunar occultations and long baseline interferometry. We derive an average value of $19.96 \pm 0.03$ milliarcsec for the uniform disk diameter. The corresponding limb-darkened value is $20.58 \pm 0.03$ milliarcsec, or $44.2 \pm 0.9\,R_\odot$. We discuss this result, in connection with previous determinations and with possible problems that may affect such measurements.

## 1. Introduction

The first-magnitude star Aldebaran ($\alpha$ Tau, HD 29139, SAO 94027) is one of the most widely known objects in the sky. With a spectral type K5III, its deep orange color makes it one of the most easily recognizable stars from the northern and most of the souther emisphere. Its late spectral type makes it one of the brightest near-infrared stars in the sky as well, with $K \approx -3$ mag. At a distance of just $19.96 \pm 0.38$ pc, as measured by the Hipparcos satellite, its angular size is among the largest of all stars. Furthermore, this star is favorably located on the apparent Moon's orbit, and is therefore subject to relatively frequent lunar occultation events.

Because of these fortunate characteristics, Aldebaran has been a prime target for high angular resolution investigations, both by the technique of lunar occultations (LO) and by long-baseline interferometry (LBI), at a large number of wavelengths in the visible and near-IR. A total of about 60 measurements are listed in the CHARM Catalogue (Richichi & Percheron 2002), according to which Aldebaran is the star with the highest number of direct diameter determinations. Since Aldebaran is non-variable to a high degree of accuracy (photometric variations are known at a level $\Delta V \lesssim 0.01$ mag and will be discussed in Sect. 4.2), its angular diameter is generally assumed to be constant. With a well-measurable fringe contrast on baselines up to about 30–100 m in the near and mid-IR, Aldebaran is naturally a primary calibrator for several interferometers in the world. Consequently, Aldebaran has also represented an important building block for effective temperature calibrations, such as those of Di Benedetto & Rabbia (1987) and Perrin et al. (1998).

Here, we present new high-precision measurements by LO and LBI. The results of the two techniques are in good agreement, and are used to determine an accurate angular diameter in the near-IR under the hypothesis of a uniform disk. Subsequently, we consider a large set of previous measurements available for Aldebaran and we discuss the effects of limb-darkening and possible asymmetries and diameter variations.

This paper follows in a series devoted to angular diameter measurements of late-type stars obtained by the method of LO (see Richichi & Calamai 2003, and references therein). The series has dealt mainly with cool giants, some of which have physical characteristics very similar to those of Aldebaran. Nevertheless, we have decided to dedicate a separate paper to this particular star. This is warranted by Aldebaran's importance, and because in this case our LO measurements are complemented by LBI data.

---

* Based on observations collected at TIRGO (Gornergrat, Switzerland). TIRGO is operated by CNR – CAISMI Arcetri, Italy.

**Table 1.** Summary of the occultation observations.

| (1) Date UT | (2) Event | (3) PA ° | (4) D " | (5) Δt ms | (6) τ ms | (7) λ μm |
|---|---|---|---|---|---|---|
| 05-02-98 | D | 352 | 21 | 1.42 | 1.01 | 2.2 |
| 06-11-98 | D | 95  | 14 | 2.42 | 2.00 | 3.6 |
| 06-11-98 | R | 242 | 28 | 2.42 | 2.00 | 3.6 |

**Table 2.** Summary of the interferometric observations of Aldebaran in the $K$ band.

| UT | N. Scans | Baseline Length (m) | Baseline PA (°) | Visibility $V^2$ | Visibility $\Delta V^2$ |
|---|---|---|---|---|---|
| \multicolumn{6}{c}{12-Jan.-2002} | | | | | |
| 02:14:39 | 426 | 15.898 | 73.9 | 0.2369 | 0.0027 |
| 02:21:17 | 356 | 15.943 | 73.4 | 0.2418 | 0.0033 |
| 02:25:55 | 176 | 15.971 | 73.1 | 0.2649 | 0.0012 |
| \multicolumn{6}{c}{2-Oct.-2003} | | | | | |
| 06:50:24 | 494 | 12.876 | 80.5 | 0.4288 | 0.0019 |
| 06:56:30 | 494 | 13.114 | 80.2 | 0.4130 | 0.0022 |

## 2. Lunar occultation observations and data analysis

The latest LO series of Aldebaran occurred from 1996 to 2000, while the next series will begin in 2015. We recorded several events in the series using various instruments and telescopes. Table 1 summarizes the details of the observations. The format is similar to that used in previous papers in our series on LO angular diameters, and only slightly modified since here the source name and telescope code are always the same. Column (1) shows the UT date of observation. The event type is mentioned in Col. (2), where D and R stand for disappearance and reappearance, respectively. In Cols. (3) through (6) we list the predicted position angle (PA, measured from North to East) of occultation, the aperture of the photometer, the sampling time of the lightcurve and the integration time of each data point, respectively.

All events were observed with the TIRGO 1.5 m telescope using the facility fast photometer, details of which can be found in Richichi et al. (1997) and references therein. We used circular-variable filters (CVF), at 2.22 μm for the first event and at 3.55 μm for the other two, as listed in Col. (7). The CVF have $\lambda/\Delta\lambda \approx 70$.

In addition to the LO detailed in Table 1, we also observed an event on March 14, 1997 from the Calar Alto (Spain) observatory and from the Arcetri (Florence, Italy) institute. At Calar Alto, it was possible to record the disappearance in dispersed light in the H and K filters using the 3.5 m telescope. Although this represented an interesting first-timer for the near-IR, the time resolution of this observation was marginally sufficient to resolve the diameter of Aldebaran. In Arcetri, a small 35 cm telescope was used and the quality of the data, also considering the unfavorable location, was not very good due to strong scintillation. We do not include the results of these observations in this paper.

As customary in our series of papers on LO results, the data analysis was carried out by means of a code based on the least-squares method (LSM) and including corrections for biases due to the finite time response of the instrument, atmospheric scintillation, and pick-up frequencies. The LSM code is well suited for cases in which a model can be provided for the source, and can be characterized by a few parameters. However, a model-independent approach is needed in cases in which it is required to investigate the presence and structure of circumstellar emission and/or asymmetries of the brightness profile. For this, we have used the CAL method (Richichi 1989), which applies iteratively Lucy-Richardson deconvolution algorithm and converges to the maximum-likelihood solution.

It is interesting to note that the event of November 6, 1998 was recorded not only on the dark lunar limb, but also on the bright limb. This is a very rare combination, which could be achieved thanks to the brightness of the star, the strong reduction in intensity due to the CVF, and the very stable sky conditions. Nevertheless, the data are dominated by the background of the bright limb, and a careful analysis was required in particular to remove low-frequency fluctuations due to brightness changes induced by seeing and telescope vibrations.

The two dark-limb measurements obtained on February 5 and November 6, 1998 are of very good quality, with a signal-to-noise ratio (SNR) of 136 and 174, respectively. In particular, the first of the two events occurred with a predicted contact angle (CA) of 276°, i.e. almost grazing (the contact angle is measured between the perpendicular to the point of occultation on the lunar limb and the direction of lunar motion). This was a fortunate situation, since the rate with which the lunar limb scanned the source is then be much reduced compared to central occultations. At the same time, there was sufficient margin to avoid the influence of local limb irregularities which often affect grazing occultations. During the data analysis, the fitted rate of the event was 0.1728 m/s, indicating a local slope of 21°. The actual PA and CA resulted in 331° and 255°, safely outside the conditions of a real graze and at the same time with a relatively slow limb rate. The actual sampling of our lightcurve resulted in 0.135 mas per data point.

## 3. Interferometric observations and data analysis

The interferometric (LBI) observations have been extracted from the public data releases of the ESO Very Large Telescope Interferometer (VLTI) commissioning campaign (Schoeller et al. 2003) using the VINCI test instrument and the 35 cm siderostats. Aldebaran was observed on three different dates, two of which were of good quality and are detailed in Table 2. Observations recorded on 12 July, 2003 are also available, but they were not deemed of sufficient quality for our purposes mainly due to changes of atmospheric transfer function during the night.

Details of the VINCI data and their analysis can be found in Ballester et al. (2002) and Kervella et al. (2004). In summary, the VINCI instrument is a fiber-based beam combiner in

**Table 3.** Calibrators used in the VLTI observations.

| Star | $\phi, \Delta\phi$ (mas) | Sp. T. | Ref. (K) | # Obs. /Scans |
|---|---|---|---|---|
| 12-Jan.-2002 | | | | |
| $\alpha$ Cet | 11.69 ± 0.69 | M1.5III | 1 | 3 / 1310 |
| $\sigma$ Pup | 6.66 ± 0.10 | K5III | 2 | 2 / 884 |
| 1 Pup | 3.80 ± 0.40 | K5III | 2 | 3 / 741 |
| $\zeta$ Hya | 3.10 ± 0.20 | G9II-III | 2 | 2 / 450 |
| $\alpha$ Hya | 9.31 ± 0.16 | K3II-III | 1,3 | 4 / 649 |
| $\alpha$ Ant | 3.60 ± 0.30 | K4III | 2 | 2 / 501 |
| $\delta$ Crt | 3.10 ± 0.20 | K0III | 2 | 2 / 541 |
| $\epsilon$ Crv | 4.99 ± 0.23 | K2III | 1 | 3 / 1303 |
| 2-Oct.-2003 | | | | |
| HR 7092 | 2.81 ± 0.03 | M0III | 4 | 4 / 941 |
| $\eta$ Cet | 3.35 ± 0.04 | K1.5III | 4 | 4 / 1816 |
| $\alpha$ Cet | 11.69 ± 0.69 | M1.5IIIa | 1 | 4 / 1914 |
| $\eta$ Eri | 2.50 ± 0.12 | K1III | 2 | 3 / 1181 |
| 31 Ori | 3.56 ± 0.06 | K5III | 4 | 6 / 2124 |

References: 1) CHARM Catalogue (Richichi & Percheron 2002);
2) VLTI/VINCI Calibrators Catalogue (Percheron et al. 2003);
3) VLTI measurement (ESO Press Release 06/2001); 4) Bordé et al. (2002).

which the fringes are sampled by temporal variations of the optical path difference introduced by a piezoelectric mirror. The monomode fibers ensure a high quality of interference, albeit at the expense of efficiency in flux transmission. The flux fluctuations at the point of injection in the fibers have to be accurately monitored and are taken into account in the final processing.

Each VINCI observation consists of a number of scans of the fringes, constituting one so-called batch. In the data analysis, each scan as well as the overall batch are subject to a number of filters which are used to flag data which are invalid or of insufficient quality. In Tables 2–3 we report the total number of accepted scans for each observation.

The data are analyzed by means of an automated pipeline, whose output is uncalibrated visibilities. We note that the pipeline uses two different mathematical engines based on Fourier and wavelet transforms (see Kervella et al. 2004). The difference in the calibrated results between the two methods was consistent within the errorbars and we have adopted those based on the classical Fourier transform analysis.

As for all interferometric observations, a critical role is played by the calibrator stars, i.e. stars with a presumably well-known angular diameter which are used to calibrate the interferometer response, or so-called transfer function. Dividing the observed visibility of the science target by the interferometric transfer function yields the calibrated visibility. This latter is the quantity actually used to compute the angular diameter. The results for Aldebaran are listed in Table 2.

A calibrator should be a "normal" star, i.e. neither variable nor with an extreme spectral type. Also, it should be a single star, or any companion should be sufficiently distant or faint to avoid contributions to the visibility. At the VLTI, a list of calibrators which satisfy the above requirements has been created for the purpose of VINCI observations. It is largely based on the CHARM catalogue (Richichi & Percheron 2002), which includes most of the high angular resolution measurements appeared in the literature until 2001, as well as some indirect estimates. Also used is a catalogue of selected stars compiled Bordé et al. (2002), which includes objects with stringent accuracy requirements.

The observations of Aldebaran were done in close sequence with the calibrators. In Table 3 we list all the calibrators observed during each night, along with the number of observations (batches) and total number of valid scans. The diameters listed in Table 3 are derived from the references listed. When more than one reference is present, a weighted average has been performed. In the case of $\alpha$ Cet, where several measurements are present but they differ significantly in wavelength, we have opted for a simple average. For the sources present in the Bell & Gustafsson (1989) list, namely $\alpha$ Hya and $\epsilon$ Crv, we have assumed a 100 K uncertainty on the effective temperatures and derived the error on the diameters accordingly. The transfer functions for each night were computed on the basis of the average values of the calibrator visibilities. Small changes in the transfer function during the nights can be noticed, but it is difficult to correlate them clearly with time or with the position of the calibrator in the sky. We also checked for changes in the transfer function against the seeing values recorded at the Paranal site, without finding any obvious correlation. Since the changes are of the same order as the uncertainties due to the errors on the calibrators diameters, we have decided to compute a single transfer function value for each night, after editing and discarding problematic measurements. By this, we refer to measurements flagged by the data reduction pipeline according to the criteria described in Kervella et al. (2004) and Richichi & Percheron (2004).

In order to make proper use of both the Aldebaran and the calibrator interferometric data, it is necessary to evaluate correctly two additional quantities, namely the actual projected baseline and the effective wavelength of the observation. For the former, we have adopted the value of the projected baseline computed at the beginning of the scans. The measurement of a few hundred scans is executed for about two minutes, during which the baseline changes are relatively small. For the effective wavelength, we have followed a procedure described in an ESO technical document (Davis & Richichi 2003), which takes into account several instrumental effects as well as the spectral energy distribution of the source, defined as a blackbody with the effective temperature of the star. We have considered the uncertainties associated with both projected baseline and effective wavelength in our computations, and we have concluded that they are negligible with respect to the errors on the visibility measurements.

## 4. Results and discussion

Our angular diameter results are listed in Table 4. They have been obtained under the hypothesis of a uniform disk (UD) for both the LO and LBI observations, by minimizing the $\chi^2$ of the fit. Errors were estimated by the standard method of a unitary variation on the normalized $\chi^2$. One first consideration is the similar level of accuracy achieved by LO and LBI. With the exclusion of the LO measurement obtained on the bright limb, which necessarily yielded a lower accuracy, all other LO and LBI measurements have errors in the 0.02 to 0.15 mas range. A second important comment is that there is a general agreement between the measurements, although the scatter is significantly larger than the formal errors. In the same table we present also the weighted averages obtained by combining the LO and LBI measurements separately, and by combining the final LO and LBI results together.

**Table 4.** Individual angular diameter determinations for Aldebaran and combined values.

| Lunar Occultations | | Interferometry | |
|---|---|---|---|
| Obs. | $\phi_{UD}$ | Obs. | $\phi_{UD}$ |
| 05-02-98 | 19.85 ± 0.03 | 12-01-02 | 20.36 ± 0.13 |
| 06-11-98 | 20.18 ± 0.25 | 02-10-03 | 19.92 ± 0.05 |
| 06-11-98 | 21.30 ± 0.11 | | |
| Weighted averages | | | |
| 19.95 ± 0.03 | | 19.98 ± 0.05 | |
| Adopted uniform-disk angular diameter | | | |
| 19.96 ± 0.03 | | | |
| Adopted limb-darkened angular diameter | | | |
| 20.58 ± 0.03 | | | |

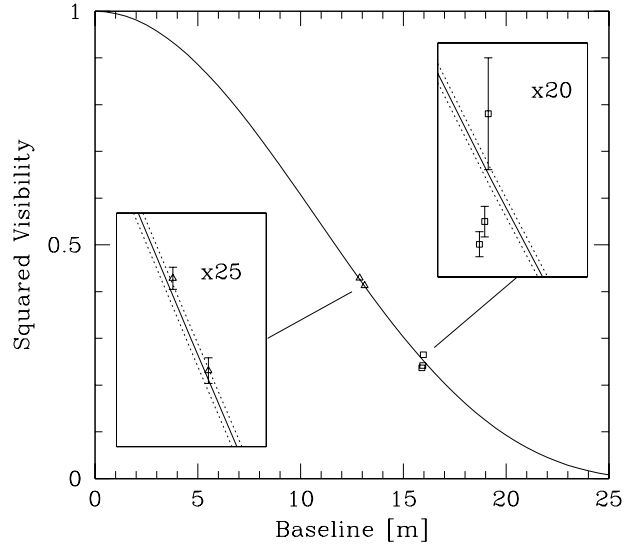

**Fig. 1.** Calibrated interferometric measurements for Aldebaran obtained on 12 January 2002 (squares) and 2 October 2003 (triangles). The solid line is a fit by the combined limb-darkened angular diameter values listed in Table 4. The insets show the data with large magnification, including the error bars and the curves corresponding to a change of one standard deviation in the angular diameter (dotted lines).

Note that a slightly different result is obtained if one performs a unique fit of both LBI runs combined: this would be the preferred approach for a uniform, perfectly circular disk without variability. In our case, we have preferred to keep the runs separated to highlight the scatter intrinsic in LBI results and in order to follow an approach similar to that used for LO results.

With this approach, the final uniform-disk (UD) angular diameter value is 19.89 ± 0.02 mas. The fit is shown in Fig. 1. The limb-darkened (LD) diameter is the preferred choice if one wants to compare results with model predictions and to derive effective temperatures, however this quantity is difficult to measure by LO and LBI, requiring extremely good SNR and observations beyond the first visibility minimum, respectively. (Richichi & Lisi 1990; Wittkowski et al. 2004). The difference between UD and LD diameters in the $K$-band for late-type non-Mira giants are predicted to be at the level of few percent (Hofmann & Scholz 1998). Computations of limb-darkened LO diameters using a grid of numerical center-to-limb variation models were reported by Richichi et al. (1999), confirming this prediction.

Courtesy of M. Wittkowski, a computation for the specific case of Aldebaran has been made using a linear parametrization of the limb-darkening based on Kurucz models (see also Wittkowski et al. 2004). The result is a ratio of 0.97 between UD and LD. This can be used to derive the effective temperature of Aldebaran, from its bolometric flux. Estimates of the bolometric flux have been given by Di Benedetto & Rabbia (1987), and more recently by Mozurkewich et al. (2003). The two estimates are in good agreement and we adopt their average and the error from the first of the two: $F = (33.57 \pm 1.35) \times 10^{-13}$ W cm$^{-2}$. Thus we obtain $T_{\rm eff} = 3934 \pm 41$ K.

### 4.1. Comparison with previous results

Placed on the zodiacal belt and having one of the largest angular diameters among all bright stars, Aldebaran probably has the most extensive list of angular diameter determinations. The CHARM Catalogue (Richichi & Percheron 2002), which includes references up to mid-2001, lists 46 independent measurements by LO, 7 by LBI and 3 indirect determinations. To this, one should add a recent LBI determination by Mozurkewich et al. (2003).

We have plotted those results having a formal error of 2.0 mas or less in Fig. 2, as a function of the wavelength of observation. One should note that we have made no attempt to convert these values to a common angular diameter convention, either UD or LD. This might introduce a small bias, which however is negligible when compared to the scatter of the individual measurements. Similarly, we have not investigated in detail the possibility that some measurements were obtained in bandpasses encompassing spectral features. We do not expect these effects, which would be limited to some visible bandpasses, to be very prominent. For example, Quirrenbach et al. (1993) have determined the angular diameter of 12 late-type giant and supergiant stars in the continuum and in a strong TiO absorption band at 712 nm, and among these Aldebaran was one of the three stars for which no difference was found. They also did not find a significant difference between the angular diameter in the visible and in the near-infrared. This is confirmed in Fig. 2 which, under the given assumptions, does not show any significant trend with wavelength. We have verified this after discarding some measurements with errorbars larger than a given threshold. As a compromise between the number of available measurements and the need to keep a significant number of good ones, this threshold was chosen at 0.4 mas. A linear fit to the best measurements did not result in a significative slope as a function of wavelength. Using the same selection criterion, the remaining 17 best results yield a weighted average of 19.87±0.05 mas. This is in agreement with our UD determination listed in Table 4, and consistent with the fact that most literature values are indeed computed under the UD assumption.

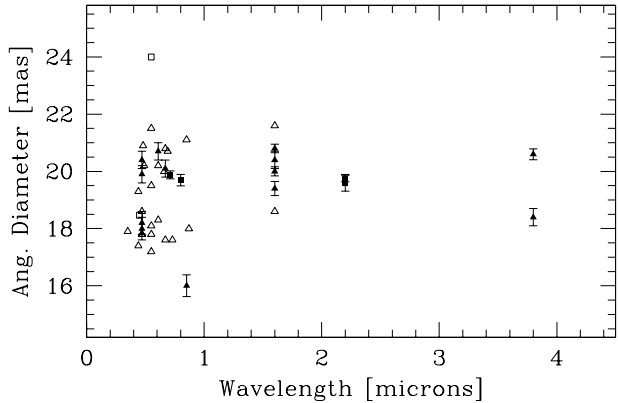

**Fig. 2.** Previous determinations of the angular diameter of Aldebaran with their error bars, as extracted from the available literature. LO and LBI measurements are shown as triangles and squares, respectively. Solid symbols mark measurements with errors of 0.4 mas or less, open symbols those with errors between 0.4 and 2.0 mas. Measurements with errors larger than 2.0 mas also exist, but are not shown here. For clarity, errorbars are shown only for the best measurements. Our result, as listed in Table 4, is in excellent agreement with the weighted average of these latter.

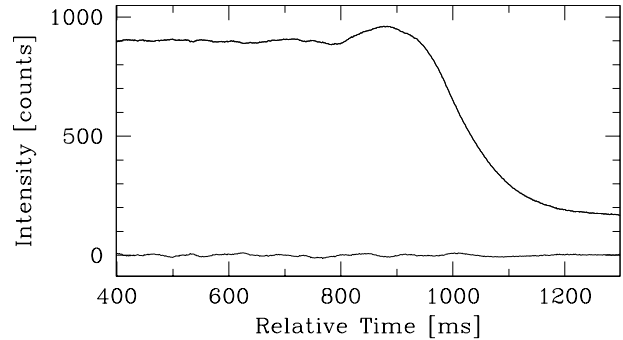

**Fig. 3.** The occultation trace of Aldebaran of February 5, 1998. The fit by a uniform circular disk was quite accurate, yielding the value reported in Table 4. The fit itself is not overplotted since it would be hard to tell from the data. Instead, we show the fit residuals in the lower curve.

Following a large number of measurements collected during the previous LO series of Aldebaran, in the years 1978 through 1980, other authors have attempted to discuss the angular diameter of Aldebaran and derive average values. Evans et al. (1980) published a critical analysis of a number of LO measurements obtained in the visible range by themselves as well as by other authors. After retaining only those measurements satisfying their criteria, the authors obtained an average diameter of 19.9 ± 0.3 mas, remarkably in agreement with our present determination. They did not find any evidence for variability with time, nor for a dependence of the diameter with wavelength in the visible range.

Ridgway et al. (1982) analyzed a set of original measurements in the range 0.4 to 3.8 $\mu$m, as well as others (some of which already included in the analysis by Evans et al. 1980). They obtained a limb-darkened diameter of 20.88 ± 0.10 mas, and additionally they strongly suggested the existence of a wavelength-dependent variation of the angular diameter.

White & Kreidl (1984) also analyzed a mix of own and literature results, again restricted to the visible in the range 0.4 to 0.9 $\mu$m, and obtained an average value of 20.45 ± 0.46 mas. White & Kreidl did not speculate on the possible wavelength dependence, but they argued that the data indicated some amount of limb darkening. They were not able to derive a single value of the limb darkening coefficient.

### 4.2. Aldebaran's elusive angular diameter

The differences between the results obtained by various authors, even after proper averaging of several determinations, is certainly puzzling. We do not have a single clear-cut explanation for this, however we note several factors which, together or separately, might account for the observed discrepancies.

Starting with LO, we note that in spite of its exceptional brightness, Aldebaran is a very difficult source to measure by this technique. The lunar limb acts as a diffracting edge, and the resulting diffraction fringes provide an excellent means to determine the angular diameter of the occulted source. However, when the source size increases, the diffraction fringes diminish progressively in contrast: this can also be seen as the transition from diffraction to classical geometrical optics. In this latter regime, the source size is related to the time required for the source to be occulted, and to the apparent speed of the lunar limb. This quantity cannot be precisely determined, since local slopes can introduce quite large uncertainties depending also on the contact angle of the event. The threshold between the diffraction and the geometrical optics regimes takes place for angular diameters $\phi_t \approx (\lambda/D)^{\frac{1}{2}}$, with $\lambda$ and $D$ being the wavelength of observation and the distance to the Moon, respectively. At 2.2 $\mu$m, $\phi_t \approx 16$ mas: therefore, Aldebaran LO traces show almost no diffraction fringes in the near-IR, as seen also in Fig. 3. The situation is in principle more favorable in the visual range, but in practice occultation traces at shorter wavelenghts are severely affected by the increased lunar background and, especially on smaller telescopes, by scintillation. An inspection of the occultation traces on which the papers mentioned above are based can quickly convince the reader of these arguments. Scintillation is a particularly adverse effect, which is prominent especially for short wavelengths and/or small telescopes, but that can affect any measurement obtained under less than optimal atmospheric conditions. Interestingly, scintillation will always provide a bias towards larger-than-real angular diameters (Knoechel & von der Heide 1978). In our series of papers we have consistently included scintillation in our data analysis when required (see Richichi et al. 1992, for a formal description), however this was generally not the case for the other literature on LO results, and in particular for the other LO-based papers mentioned in Sect. 4.1.

Concerning the LBI measurements, although there is a significant scatter in the results, this is in general less than for LO. One exception is the 24 mas result by Currie et al. (1974), which strictly speaking was obtained using amplitude interferometry on a single telescope and in any case was assigned a total uncertainty of 5 mas by the authors, sufficient to be consistent with the main trend of determinations. Indeed, Aldebaran is a much more favourable target for LBI than it is for LO, and the main limitation is set by the accuracy of calibration. It is regrettable however that no measurements are available on baselines sufficiently long to explore the visibility curve beyond the first minimum, and investigate in detail second-order effects such as limb-darkening, and expecially to probe the presence of surface structure.

Having at our disposal what could be considered to be the best LO and LBI data, and sophisticated analysis tools, we would be led to argue that our determination should represent the best available value for the angular diameter of Aldebaran. However, the scatter in the results of Table 4 should not be overlooked. Both LO and LBI provided individual measurements which differ significantly more than the associated errors. A first explanation is of course that there are systematic errors in the measurements. In the case of LO this could be due to scintillation, inaccuracy of the effective bandpass, insufficient knowledge of the time response of the instrument. We note however that such effects are more prominent for LO measurements of stars with a smaller diameter, while in the case of Aldebaran it is not expected that their importance would be crucial. In the case of LBI, possible causes of systematic errors are in the inaccuracies of calibrators and in the effect of variations in the transfer function of the instrument and – more significantly – of the atmosphere. At least some of the errors are expected to tend to zero with a large number of measurements, but in our case we only have a small data set.

However, the scatter of the angular diameter results available for Aldebaran is remarkably larger than what is normally experienced. For example, the VLTI is capable of repeated measurements on calibrator stars which have an accuracy much smaller than the difference in the LBI results shown in Fig. 4. It is therefore interesting to consider other intrinsic reasons, such as variability in the diameter, or non-spherical symmetry, or a non-homogeneous stellar disc.

Variability among late-type giants has been generally accepted as the norm for late M stars. But only recently it has been recognized that K giants are a new class of photometric and radial velocity variables, although with much lower amplitudes. As far as radial velocity (RV) is concerned, the recent survey of Setiawan et al. (2003, 2004) has shown the ubiquity of RV variability among K giants, with indications that it increases with the luminosity of the stars. Concerning photometric variability, in a survey by Henry et al. (2000) 50% of the K giants were found to be variable. The main causes are stellar pulsation, and surface features such as spots which produce rotational modulations of the flux.

In the case of Aldebaran a small amount of variability is known, although with some ambiguities in its amplitude and period. It is reported has having a variation of 0.2 mag in the GCVS (Kholopov et al. 1988). A study over three years by Krisciunas (1992) found the $V$ magnitude to be stable within 0.004 mag. A study by Wasatonic & Guinan (1997), in contrast, found a variation of about 0.018 mag over about six months, and suggested a periodicity of 91 days. Henry et al. (2000) found a dispersion of 0.0069 mag over 500 days, with no clear indication of periodicity. By extent and number of observations this is probably the most convincing study, although long-term variations of the characteristics of variability cannot be excluded.

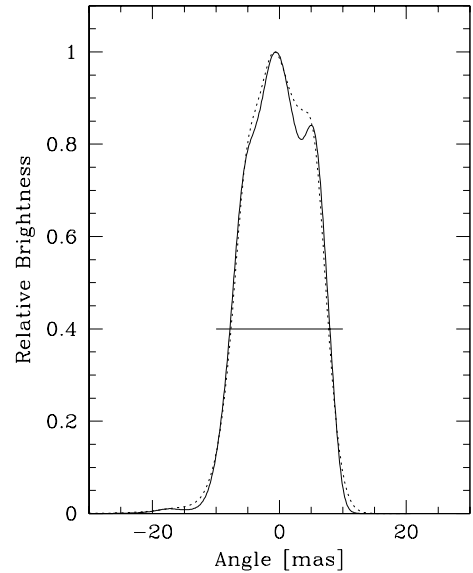

**Fig. 4.** Brightness profile of Aldebaran derived from the data of February 5, 1998 using a model-independent analysis. The dashed line refers to the profile after 1000 iterations, the solid line at convergence after about 4000 iterations (see text). The horizontal segment marks the size of the equivalent uniform disk diameter (see Table 4).

If the change in brightness was due to a pure oscillation mode, it could be considered to be at a fixed photospheric temperature, and the corresponding angular diameter variation would be about 0.83%, or about 0.17 mas peak to peak. If such pulsation is accompanied by a change in effective temperature, as in a thermal cycle in the photosphere, it is expected that the higher luminosity would be observed when the diameter is smaller. Therefore, the value of 0.17 mas could be an upper limit. Nevertheless, it is interesting to note that it is not negligible at the level of the uncertainties that we are dealing with in the case of the diameter measurements of Aldebaran, and its effect on the scatter of individual measurements cannot be completely excluded.

It is also interesting to mention that periodicities have been observed in the radial velocity and in the bisector profile of the TiI 6304 Å line (Hatzes & Cochran 1998a, 1998b), but with different values of $654^d$ and $49^d\!.9$, respectively. Since rotational periods in this class of stars are typically of the order of a few $100^d$, it seems plausible that the short-period variations could be caused by pulsation, and in particular non-radial modes would be necessary to explain the bisector profiles. The long-period variations could be caused by a nearby low-mass companion, and the above mentioned authors speculated on the presence of a 11 $M_J$ exoplanet that would induce astrometric perturbations of 0.8 mas.

Another important possible cause of both variability and RV periodicities could be the presence of surface spots. Naturally, this could also constitute an important factor in the determination of the angular diameter and in the scatter of the measurements. Hatzes & Cochran (1998b) suggested that the observed RV amplitude would be consistent with surface spots having 1000 K $\Delta T$ causing photometric fluctuations of about 0.14 mag, marginally consistent with the observations.

At present, it seems difficult to confirm observationally these various hypotheses. Concerning the reflex motion induced by a possible low-mass companion, future investigations by the VLTI using its narrow-angle phase-referenced astrometric mode (PRIMA, Paresce et al. 2003) will be able to provide the required level of accuracy. It is difficult to test the hypothesis of surface spots by LBI measurements. However, LO data can be used to produce model-independent brightness profiles (Richichi 1989). In Fig. 3 we show the data of the February 5, 1998 event together with the best fit by a uniform disk model. Although the fit is in good agreement with the data, some discrepancies can be seen. In order to obtain a model-independent fit, we used the CAL method, the result of which is shown in Fig. 4. In this case, the result is more convincing and statistically significant: the reduced $\chi^2$ improved from 1.6 to 0.9.

We conclude that the observed LO data would be consistent with the presence of surface features on Aldebaran. We stress that such claims have been made before for other stars (see for example Richichi & Lisi 1990). It is however difficult to demonstrate this hypothesis convincingly, also in view of possible minor perturbations due to local irregularities in the lunar limb. The contact angle for the February 5, 1998 event was very large, i.e. not under grazing conditions which would amplify the effect of such irregularities. Nevertheless, the scale subtended by Aldebaran at the lunar limb is so large that small changes in slope can occur. At the first order such slope changes would result in a variation of the fringe frequency. These were not observed, although our measurement is not very sensitive to this parameter.

## 5. Conclusions

We have presented a set of new accurate measurements of Aldebaran by means of lunar occultations and long-baseline interferometry, which we have analyzed and combined to derive an accurate angular diameter. The result is 19.96 ± 0.03 mas and 20.58 ± 0.03 mas for the uniform and limb-darkened disk cases, respectively. Correspondingly, we deduce its linear radius and effective temperature.

In spite of the very high formal accuracy of this result, our measurements show an intrinsic scatter which could be indicative of either variability, or of departure from the simple stellar disk hypothesis. The fact that also the large number of measurements available from the literature shows a significant scatter, although with varying levels of accuracy, seems to lend additional weight to this hypothesis.

Evidence exists, from photometric as well as spectroscopic studies, that Aldebaran could be subject to small changes in angular diameters and/or have a small but significant amount of surface spots. Both the amplitudes and the characteristic periods of these phenomena would be consistent with the scatter observed in the results of high angular resolution measurements. We have also noted that the brightness profile of Aldebaran's photosphere, reconstructed from a high-SNR lunar occultation trace, would be consistent with a non-uniform stellar disc, i.e. with the presence of spots. The additional hypothesis of a low-mass companion cannot be proven at present, but will be tested with the forthcoming PRIMA facility of the VLTI. Another explanation to be considered is of course the effect of systematic errors. This hypothesis is difficult to test, and would justify a more extended campaign of measurements.

Thus, it appears that the angular diameter of Aldebaran can be considered reliable to a precision level of about 0.5%. This permits its use for effective temperature calibrations to about 50 K and for interferometric visibility calibrations to about 1% (assuming a 10–20 m baseline in the near-IR). However, a higher level of accuracy can only be achieved by a combined study by interferometry and spectroscopy. In particular, simultaneous observations by these techniques at various phases during a time comparable with the estimated rotation period of 654 days would be a challenging, but highly rewarding project.

*Acknowledgements.* This research has made use of the *Simbad* database, operated at the Centre de Donnés Astronomiques de Strasbourg (CDS), and of NASA's Astrophysics Data System Bibliographic Services (ADS). VR's stay at ESO was partly financed by the italian National Institute for astrophysics (INAF), under grant No. 0330909. We are grateful to the many ESO colleagues who have contributed to the large effort of the VLTI/VINCI commissioning observations and data analysis, and in particular to I. Percheron and M. Wittkowski. We are also grateful for useful discussions with D. Baade and L. Pasquini.